# Formation of High Pressure Gradients at the Free Surface of a Liquid Dielectric in a Tangential Electric Field

Evgeny A. Kochurin
Institute of Electrophysics
Ural Branch of the Russian Academy of Sciences
Yekaterinburg, Russia
e-mail: kochurin@iep.uran.ru

*Abstract*—**Nonlinear dynamics of the free surface of an ideal incompressible non-conducting fluid with high dielectric constant subjected by strong horizontal electric field is simulated on the base of the method of conformal transformations. It is demonstrated that interaction of counter-propagating waves leads to formation of regions with steep wave front at the fluid surface; angles of the boundary inclination tend to π/2, and the curvature of surface extremely increases. A significant concentration of the energy of the system occurs at these points. From the physical point of view, the appearance of these singularities corresponds to formation of regions at the fluid surface where pressure exerted by electric field undergoes a discontinuity and dynamical pressure increases almost an order of magnitude.**

## I. INTRODUCTION

It is well known that an external electric field directed tangentially to the unperturbed contact or free surfaces of dielectric liquids has a stabilizing effect on the boundary motion [1]. On the other hand, the perpendicular field results in aperiodic growth of the boundary perturbations (electro-hydrodynamic instability) [2]. At the present time, the processes of nonlinear structures formation on the interface of fluids in a vertical electric field are studied sufficiently (see [3] and its references list). Thus, the interest to study the nonlinear dynamics of fluids interfaces under the action of an electric field is caused by the possibility of controlling and suppressing of boundary unsteadiness by means of the applied field. The stabilization processes of Rayleigh-Taylor and Kelvin-Helmholtz instabilities by the external electric field are investigated in [4-5] and [6-7], correspondingly. In general, the stabilization of hydrodynamic instabilities is a complex problem; for the solution of which, it is necessary to understand how the liquid boundary behaves itself in a strong horizontal electric field without destabilizing factors.

Nonlinear dynamics of the free surface of deep non-conducting liquid with high dielectric constant in the absence of instabilities is studied in the present work. In the case of high permittivity, the normal (destabilizing) component of the field is much smaller than tangential (stabilizing) one. It is known [8] that nonlinear waves on such surface can propagate along or against the electric field direction without distortion, i.e., the equations of motion admit a wide class of exact traveling wave solutions. This result was generalized in [9] for the weakly nonlinear waves on the boundary between two immiscible dielectric fluids. It has been demonstrated in [10] that the interaction between oppositely propagating waves of arbitrary geometry is elastic (they conserve their energy and momentum) and it is relatively weak for localized waves.

The methods of computer simulation based on the dynamic conformal transforms of the region occupied by the fluid into half-plane of auxiliary variables were used in the present work. This approach was firstly suggested in [11] for the study of nonlinear surface waves under the action of gravity. The simulations show that the boundary can deform in result of collisions of the counter-propagating waves. The process of waves interaction leads to formation of regions at the fluid surface with a high energy concentration. The fluid velocity and pressure exerted by electric field undergo a jump in these singular points. The calculations have shown that the angles of the boundary inclination tend to $\pi/2$ and the curvature of surface increases unlimitedly in finite time of the system evolution.

## II. INITIAL EQUATIONS

We consider a potential flow of an incompressible ideal dielectric liquid of infinite depth with a free surface in an external uniform horizontal electric field. The boundary of the liquid in the unperturbed state is the horizontal plane $y=0$ (the $x$ axis of the Cartesian coordinate system lies in this plane and the $y$ axis is perpendicular to it). Let the function $\eta(x, t)$ specify the deviation of the boundary from the plane; i.e., the equation $y=\eta$ determines the profile of the surface. Let the electric field be directed along the $x$ axis and be $E$ in magnitude.

We consider the case where the dielectric constant of the liquid is large, $\varepsilon \gg 1$. As was shown in [8] and was used, e.g., in [10], the normal component of the electric field in the liquid in this limit is much smaller than the tangential component. This means that field lines inside the liquid are directed along a tangent to its surface. In this case, the field distribution in the liquid can be determined disregarding the field distribution above it.

Let us write the governing equations describing nonlinear evolution of the system under consideration. The velocity potential of the liquid $\Phi$ and electric field potential $\varphi$ satisfy the Laplace equations $\Delta\varphi=0$, $\Delta\Phi=0$. They should be solved together with the following conditions at the boundary and infinity:

$$\Phi_t + (\nabla\Phi)^2/2 = \varepsilon[(\nabla\varphi)^2 - E^2]/(8\pi\rho), \quad y = \eta(x,t), \quad (1)$$

$$\varphi_y - \eta_x\varphi_x = 0, \quad y = \eta(x,t),$$

$$\Phi \to 0, \quad \varphi \to -Ex, \quad y \to -\infty,$$

where $\rho$ is the density of the liquid. Time dependent Bernoulli equation (1) is written for the strong field limit (for more details, see [8, 12]), in which the motion of the boundary is determined by the electrostatic forces (capillary and gravitational forces are disregarded). The equations of motion are closed by the kinematic relation

$$\eta_t = \Phi_y - \eta_x\Phi_x, \quad y = \eta(x,t).$$

It is important to note that in the framework of these equations the propagation of surface linear waves is dispersionless. They move without distortion along the boundary of the liquid at the velocity $c = (\varepsilon E^2/4\pi\rho)^{1/2}$. For convenient further consideration, we pass to the dimensionless variables as

$$\Phi \to \lambda c\Phi, \quad \varphi \to \lambda E\varphi, \quad x \to \lambda x, \quad y \to \lambda y, \quad t \to \lambda c^{-1}t,$$

where $\lambda$ is the characteristic wavelength. In the new variables, the velocity of linear waves is unity.

By analogy with [10-11], we make the conformal transformation of the region occupied by the liquid to the parametric half-plane $-\infty < v \le 0$ and $-\infty < u < +\infty$. The auxiliary variables $u$ and $v$ in the problem under study have clear physical meaning: $u$ coincides with the field potential $\varphi$ except for the sign and the condition $v = $ const specifies the electric field lines. In the new variables, the Laplace equations for the electric field potential and velocity potential can be solved analytically. As a result, the initial problem of motion of the liquid can be reduced to the problem of motion of its free surface, which has a lower dimension of (1+1). The surface of the liquid in the new variables is specified by the parametric expressions $y=Y(u,t)$, $x=X(u,t)=u-\hat{H}Y$, where is $\hat{H}$ the Hilbert transform defined as

$$\hat{H}f = \frac{1}{\pi}\text{V.P.}\int\frac{f(u')}{u-u'}du'.$$

The relation between the functions $\eta(x, t)$ and $Y(u, t)$ is given in the implicit form $Y(u,t) = \eta(u-\hat{H}Y)$. In Fourier space the action of the Hilbert transform has the following form $\hat{H}\exp(iku)=i\text{sgn}(k)\exp(iku)$.

Let us introduce the complex functions $Z=X+iY$ and $\Omega=\Psi+i\hat{H}\Psi$, where $\Psi(u,t)$ defines the value of velocity potential at the boundary $v=0$. The functions $Z$ and $\Omega$ are analytic in the lower half-plane of the complex variable $u$. It is also convenient for the numerical analysis to pass to so-called Dyachenko variables [11]:

$$R = 1/Z_u, \qquad V = i\Omega_u/Z_u.$$

As a result, the initial equations system takes the compact and symmetric form (for more details, see [8]):

$$R_t = i(UR_u - U_uR), \qquad V_t = i(UV_u - D_uR), \qquad (2)$$

where we introduced the notations

$$U = \hat{P}(V\bar{R} + \bar{V}R), \qquad D = \hat{P}(V\bar{V} - R\bar{R}),$$

and $\hat{P} = (1+i\hat{H})/2$ is the projector (it makes any function analytic in the lower half-plane). The overline above the functions $R$ and $V$ stands for complex conjugation

It should be noted that the square root of the inverse Jacobian of the transform $J^{-1/2} = (X_u^2 + Y_u^2)^{-1/2} = |R(u,t)|$ has a sense of the absolute value of electric field strength, i.e., it defines the electrostatic pressure at the boundary $P_E=|R(u,t)|^2$. The modulus of $V(u,t)$ determines the fluid velocity on the surface, consequently, $P_V=|V(u,t)|^2$ corresponds to the density of kinetic energy or dynamic pressure at the boundary.

### III. SIMULATION RESULTS

The equations system (2) has exact solutions corresponding to surface waves of arbitrary geometry that propagate at a constant velocity without distortions along or against the external electric field, i.e., similar to linear waves. It should be noted that this situation is similar to that for the Alfven waves in an ideal fluid. The wave packets of arbitrary forms can travel nondispersively with the Alfven speed in or against the direction of the external magnetic field. In the problem under consideration, the nonlinearity greatly affects on the process of interaction between oppositely propagating waves. The fundamental question arises: what are consequences of such interaction? It is necessary to integrate the system (2) numerically for the answer to that.

To numerically solve (2), it is convenient to use the pseudo-spectral methods of calculations, i.e., all functions are approximated by finite Fourier series. So, let us consider interaction of counter-propagating periodic waves with the initial conditions: $R(u,0)=1+0.2\exp(iu)+0.015\exp(2iu)$, and $V(u,0)=-0.2i\exp(iu)+0.15i\exp(2iu)$.

Number of Fourier harmonics used in calculations presented below was equal to 16384, the spatial period of the problem was $2\pi$. The integration with respect to time was carried out by the explicit Runge-Kutta method of fourth-order accuracy, the discretization interval was $5\cdot10^{-5}$. The shape of surface (a), electric field pressure (b) and dynamic pressure (c) are shown in Fig. 1, correspondingly.

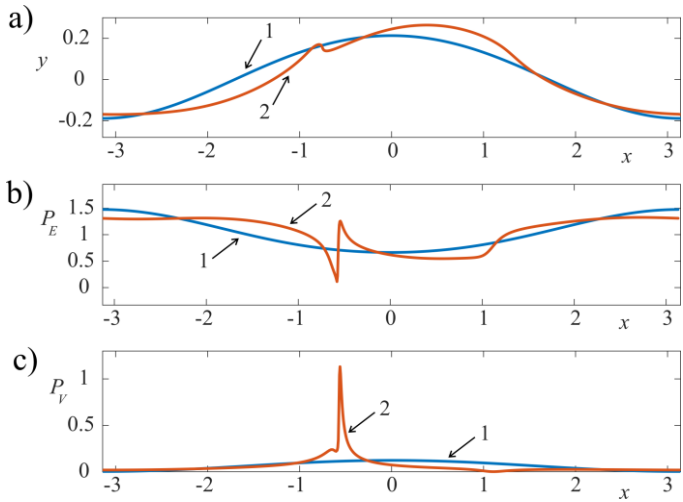

Fig. 1. The surface of liquid (a), electric field pressure (b) and dynamic pressure are shown at the initial moment (blue lines "1'') and at the end of calculation interval (red lines "2'').

It can be seen that a region with steep wave front has been formed at the end of calculation interval ($t\approx31.8$). In this point, the pressure exerted by electric field undergoes a discontinuity, and dynamic pressure increases almost an order of magnitude. The spatial-temporal evolution of electric field pressure is plotted in Fig. 2. As we can see, at the initial stages of the system evolution the quantity $P_E$ is enough smooth. In some moment the interaction of surface waves leads to formation of narrow spatial regions, where electric field changes sharply. In fact, a high concentration of kinetic energy occurs at these points. The Fig. 3 shows the evolution of dynamic pressure or, equivalently, the density of kinetic energy at the boundary. The behavior of $P_V$ is slightly different than one presented for the electric field pressure. In this case, the interaction of counter-propagating waves results in increasing of the amplitude of kinetic energy density. It means that the fluid velocity, at the points where electric field has a jump, extremely increases. It is interesting to note that magneto-hydrodynamic Alfven waves in conducting medium have the similar property; their interaction can lead to plasma acceleration [13].

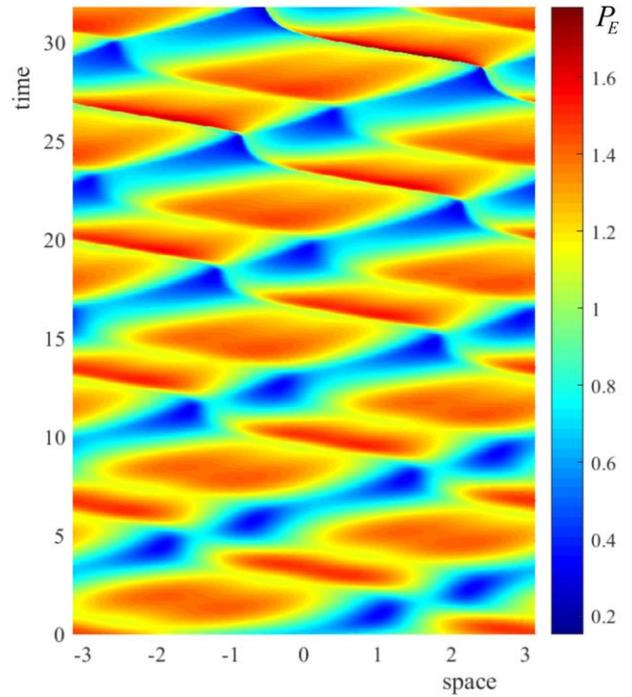

Fig. 2. The electric field pressure at the boundary of the liquid versus time.

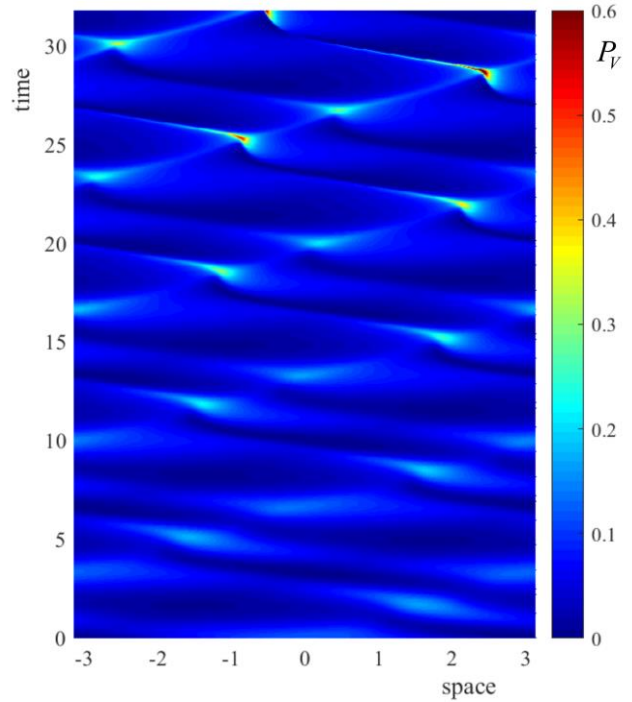

Fig. 3. The dynamic pressure at the boundary of the liquid versus time.

Accordingly to calculations, angles of the boundary inclination tend to 90º (at the initial moment it did not exceed 10º). In the same time, the curvature of surface,

$$K = \eta_{xx} / (1+\eta_x^2)^{3/2},$$

increases approximately in three orders of magnitude, i.e., unlimitedly. The Fig. 4 shows the curvature maximum versus time. It can be seen that it exponentially rises after each interaction of waves. In general, the described behavior of the surface is very similar to Riemann wave breaking, which is observed in gas and fluid dynamics [14] and leads to shock wave formation with discontinuous change in pressure and density.

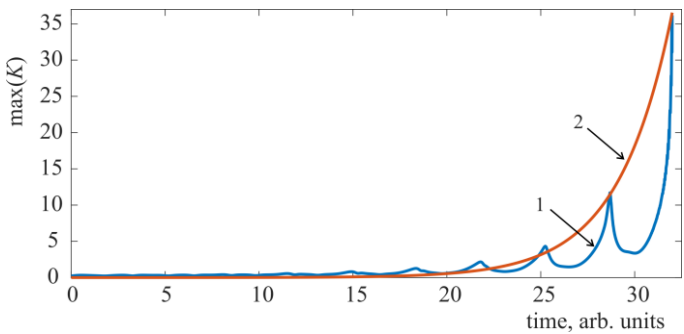

Fig. 4. The maximum of surface curvature versus time (blue line "1''"), an exponential dependence (red line "2''").

## IV. CONCLUSION

In the present work, the nonlinear dynamics of the free surface of an ideal dielectric liquid in a strong horizontal electric field is simulated in the framework of method based on dynamic conformal transforms. It has been shown that high concentration of kinetic energy occurs in the narrow spatial regions in result of the interaction between counter-propagating periodic waves. In these points, the pressure exerted by electric field undergoes a discontinuity, and dynamic pressure increases almost an order of magnitude. Numerical experiments have shown that angles of the boundary inclination tend to 90º and curvature of surface increases unlimitedly. From the physical point of view, it can mean that, from some moment in time, the effect of surface tension will play a sufficient role, and it will be necessary to take it into account.

## ACKNOWLEDGMENT

This work was supported by the state contract no. 0389-2014-0006, RFBR (projects No 16-38-60002 mol_a_dk, 16-08-00228, and 17-08-00430), by the Presidential Programs of Grants in Science (project SP-132.2016.1), and by the Presidium of the Ural Branch, Russian Academy of Sciences (project no. 15-8-2-8).